\documentclass[acmtog,balance=True]{acmart}

\copyrightyear{2026}
\acmYear{2026}
\setcopyright{cc}
\setcctype{by}
\acmConference[SIGGRAPH Conference Papers '26]{Special Interest Group on Computer Graphics and Interactive Techniques Conference Conference Papers}{July 19--23, 2026}{Los Angeles, CA, USA}
\acmBooktitle{Special Interest Group on Computer Graphics and Interactive Techniques Conference Conference Papers (SIGGRAPH Conference Papers '26), July 19--23, 2026, Los Angeles, CA, USA}
\acmDOI{10.1145/3799902.3811056}
\acmISBN{979-8-4007-2554-8/2026/07}

\usepackage{url}
\usepackage{subfig}
\usepackage{subcaption}

\usepackage{glossaries}
\newacronym{csf}{CSF}{contrast sensitivity function}
\newacronym{cpd}{cpd}{cycles per degree}
\newacronym{cpp}{cpp}{cycles per pixel}
\newacronym{hvs}{HVS}{human visual system}
\glsdisablehyper

\graphicspath{{fig/}}

\begin{document}

\title{It's Not Just a Phase: Creating Phase-Aligned Peripheral Metamers}

\author{Sophie Kerga{\ss}ner}
\email{sophie.kergassner@usi.ch}
\affiliation{\institution{Università della Svizzera italiana}\country{Switzerland}}

\author{Piotr Didyk}
\email{piotr.didyk@usi.ch}
\affiliation{\institution{Università della Svizzera italiana}\country{Switzerland}}

\begin{abstract}
Novel display technologies can deliver high-quality images across a wide field of view, creating immersive experiences. While rendering for such devices is expensive, most of the content falls into peripheral vision, where human perception differs from that in the fovea. Consequently, it is critical to understand and leverage the limitations of visual perception to enable efficient rendering. A standard approach is to exploit the reduced sensitivity to spatial details in the periphery by reducing rendering resolution, so-called foveated rendering. While this strategy avoids rendering part of the content altogether, an alternative promising direction is to replace accurate and expensive rendering with inexpensive synthesis of content that is perceptually indistinguishable from the ground-truth image. In this paper, we propose such a method for the efficient generation of an image signal that substitutes the rendering of high-frequency details. The method is grounded in findings from image statistics, which show that preserving appropriate local statistics is critical for perceived image quality. Based on this insight, we extrapolate several local image statistics from foveated content into higher spatial frequency ranges that are attenuated or omitted in the rendering process. This rich set of statistics is later used to synthesize a signal that is added to the initial rendering, boosting its perceived quality. We focus on phase information, demonstrating the importance of its alignment across space and frequencies. We calibrate and compare our method with state-of-the-art strategies, showing a significant reduction in the content that must be accurately rendered at a relatively small extra cost for synthesizing the additional signal.
\end{abstract}



\begin{teaserfigure}
\vspace{-3pt}
\hspace{-10pt}\includegraphics[width=530.3pt]{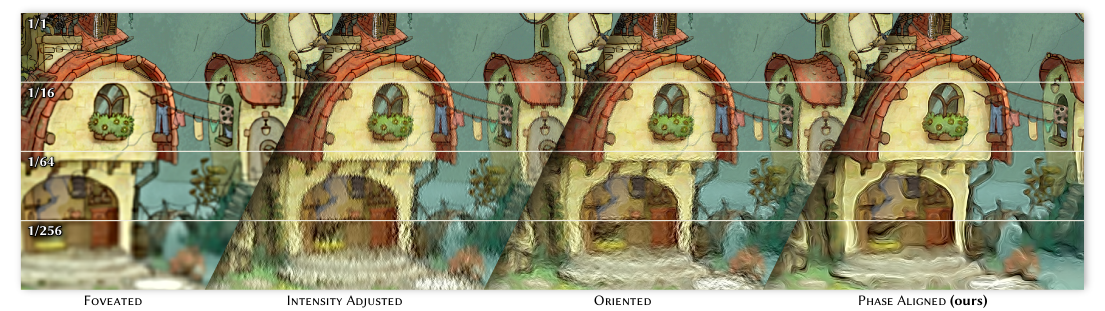}
    \setlength{\abovecaptionskip}{-3pt}
\caption{Three of our examined enhancement methods: \textsc{Intensity Adjusted}, \textsc{Oriented}, and \textsc{Phase Aligned}.
Each of them recover an increasing number of statistical properties extracted solely from the underlying foveated base image. 
The statistical properties are then used to resynthesize the missing high frequency signal.
The numbers indicate the shading rate reduction of the base image, from which we extract and extrapolate three statistical parameters: local intensity, orientation, and phase alignment across space and frequencies.
Compared to common foveation, our enhancement allows an approximately $4\times$ smaller shading rate while remaining indistinguishable to a full resolution image.
\textcopyright \thinspace Blender Studio}
\label{fig:teaser}
\Description{}
\end{teaserfigure}

\maketitle

\section{Introduction}
Despite advances in graphics hardware, rendering for wide-field-of-view immersive displays, such as virtual and augmented reality headsets, remains a significant challenge. While high spatial and temporal resolution is required for perceived quality and comfort, content complexity and realism are often compromised to achieve sufficient computational and power efficiency. Foveated rendering has emerged as an essential perception-based rendering strategy for optimizing rendering costs without sacrificing quality \cite{mohanto-2022}. The standard approaches exploit the lower sensitivity of the \gls{hvs} to spatial distortions at high spatial frequencies in the periphery and reduce the shading resolution \cite{guenter-2012a} or the geometric complexity \cite{murphy-2001} in these regions. In fact, these methods are instances of a broader class of techniques that aim to efficiently generate \textit{metamers}, i.e., images that differ physically but are perceptually indistinguishable from the original rendering. Unfortunately, despite a body of prior research, it remains unclear what constitutes a good peripheral metamer for foveated rendering, i.e., an image that is inexpensive to synthesize and metameric to the original rendering. There are likely many ways to obtain such an image, yet finding the optimal one remains an open research question.

Prior studies of peripheral vision provide valuable insight into the creation of peripheral metamers. Besides the well-known decay in sensitivity to high spatial frequencies \cite{strasburger-2011}, the \gls{hvs} remains sensitive to a range of spatial frequencies for which reproduction of other statistics, such as orientation and phase, is less critical \cite{thibos-1996a}. Other studies revealed the importance of more complex summary image statistics. They were essential for synthesizing texture metamers \cite{portilla-2000a}, modeling crowding \cite{balas-2009a}, and visual search \cite{rosenholtz-2012a}. These studies suggested that preserving higher-order statistics is critical for creating effective metamers, leading to several methods that synthesize metamers from ground-truth images \cite{fridman-2017,walton-2021a,kaplanyan-2019}. The crucial difference in the context of real-time rendering is that ground-truth information is unavailable. A viable approach to address this problem is to combine the standard foveated rendering, which reduces shading rate, with an inexpensive enhancement step that adds a signal which mimics the original image statistics. Such a strategy was previously explored by \citeauthor{patney-2016a}~\shortcite{patney-2016a}, who proposed a method to enhance contrast information, and by \citeauthor{tariq-2022a}~\shortcite{tariq-2022a}, who enhanced foveated rendering with noise of specific frequency and orientation.  Both methods explored the correlation between image content at different frequency scales, allowing the enhancement parameters to be derived directly from the rendered image.

\begin{figure}
    \includegraphics[width=\linewidth]{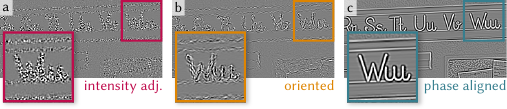}
    \setlength{\abovecaptionskip}{-7pt}
    \caption{The impact of recovering an increasing amount of local statistics: (a) solely local intensity is recovered (b) additionally, orientation is recovered (c) neighboring kernels are phase aligned.
    \textcopyright \thinspace Christophe Seux}
    \label{fig:importance-phase}
    \Description{}
\end{figure}

While prior methods proved efficient at enhancing foveated rendering by focusing on contrast and local orientation, they fail to correctly reproduce image edges, which are essential even in the far periphery \cite{morrone-1987a, morrone-1989a}. Interestingly, image edges arise from phase correlation across space and frequency (Figure~\ref{fig:importance-phase}), 
also noted in prior work as a vital image statistic for metamer creation \cite{balas-2009a,rosenholtz-2016a, freeman-2011a}. 
To address this gap, we propose a new method that enhances foveated rendering by incorporating not only local intensity and orientation, but also phase information. Our method relies solely on the foveated image as input and leverages information preserved in the low-frequency bands 
to estimate the local intensity, orientation, and phase statistics required for enhancement. We exploit the assumption that phase relationships remain coherent across frequency and space, enabling the reconstruction of consistent structures such as lines and edges \cite{rosenholtz-2016a, portilla-2000a}. Based on the estimated statistics, our method generates a signal that is added to the foveated input. In a perceptual experiment, we evaluate the importance of the considered statistics and demonstrate that additional phase incorporation enables stronger foveated rendering, leading to computational savings. Finally, we compare our technique with prior methods, confirming that, at negligible cost, it can significantly reduce the amount of information rendered in the first step.

\section{Background}

A long-standing goal in graphics and vision research is to exploit the limitations of the \gls{hvs} in order to reduce computational and display costs without introducing perceptually noticeable artifacts. This idea is commonly referred to as perceptual optimization or \textit{visual metamerism}, where two images that differ in their physical representation are nevertheless indistinguishable to a human observer.

\paragraph{Visual Metamerism}
Broadly, existing approaches can be categorized based on whether they preserve spatial image content and instead modify its presentation, or whether they directly alter the spatial structure of the image itself. The former class includes techniques that exploit perceptual insensitivities by changing global or local presentation parameters, such as color \cite{duinkharjav-2022a}, brightness and luminance \cite{surace-2025a}, or temporal resolution \cite{jindal-2021a}. These methods often yield savings primarily at the display or transmission stage. In contrast, a second class of methods directly modifies the spatial representation of image content to achieve computational savings during rendering. Prominent examples include foveated rendering and its extensions \cite{mohanto-2022}. While early frameworks reduced rendering resolution based purely on eccentricity \cite{guenter-2012a, patney-2016a}, more recent approaches incorporate masking properties of the underlying content \cite{tursun-2019a}, as well as attention-driven adaptations \cite{krajancich-2023a}.
To mitigate the resulting loss of spatial detail, a variety of recovery or enhancement techniques have been proposed,  including contrast enhancement \cite{patney-2016a}, learning-based reconstruction \cite{kaplanyan-2019}, and noise-based enhancement \cite{walton-2021a, tariq-2022a}. A closely related line of work focuses on super-resolution and upsampling. Image super-resolution methods directly predict missing details without requiring additional rendering information \cite{wang-2021a, chen-2023a, jo-2018a}. Rendering-oriented upsampling techniques take advantage of auxiliary data from the rendering pipeline to reconstruct full-resolution images from lower-resolution renderings. Examples include NVIDIA DLSS\footnote{\url{www.developer.nvidia.com/rtx/dlss}} and Meta Neural Supersampling \cite{xiao-2020a}.

\paragraph{Texture Synthesis and Image Statistics}
Image structures and therefore their recovery are essential for object recognition and scene understanding. A substantial body of work on visual metamers draws inspiration from texture synthesis, a field whose primary goal is to reproduce perceptually similar images by matching selected statistical properties \cite{heeger-1995a, portilla-2000a}. Unlike global texture synthesis, however, peripheral models often operate on local regions to reflect spatially varying visual sensitivity \cite{balas-2009a, freeman-2011a}. Creating perceptually indistinguishable metamers has therefore been closely tied to identifying which statistical image properties are critical for visual perception \cite{portilla-2000a, rosenholtz-2016a}. Among these properties, phase information has been repeatedly shown to play a central role.

\paragraph{Phase Information}
Early studies on Fourier image representations demonstrated that phase carries much of the semantically relevant information in natural images. Classic experiments showed that swapping phase spectra between images largely transfers recognizable structure \cite{piotrowski-1982a, oppenheim-1981a}, whereas swapping amplitude spectra does not, as long as it remains plausible for natural content \cite{tadmor-1993a}. Subsequent work confirmed the importance of phase for object and scene recognition, showing that increasing phase noise progressively impairs recognition performance \cite{wichmann-2006a}. Moreover, local phase coherence has been shown to be a defining cue for edges and contours \cite{morrone-1987a,gladilin-2015a}, as well as perceived image sharpness \cite{wang-2003a}. In particular, disruptions of phase coherence across space and scale contribute significantly to the perception of blur \cite{wang-2003a}. Neuroscientific and psychophysical studies further suggest that phase information is encoded early in visual processing and plays a role in shape perception \cite{pollen-1981a}. Importantly, several studies have investigated the perception of phase in peripheral vision. While some work suggests that sensitivity to absolute phase relationships decreases with eccentricity \cite{rentschler-1985}, other findings indicate that coherent phase alignment across space remains perceptually salient, even in the periphery \cite{morrone-1989a}. This distinction is critical: although observers may be less sensitive to precise phase values in terms of absolute positional encoding, they can still detect whether phase \textit{relationships} are structurally consistent \cite{morrone-1989a}. By focusing on statistical properties that govern structural coherence rather than exact spatial detail, our method aims to produce enhanced peripheral imagery that remains perceptually faithful while enabling stronger foveation and greater computational savings.

\section{Extracting and Extrapolating Image Statistics}

\begin{figure}
\vspace{-3pt}
    \includegraphics{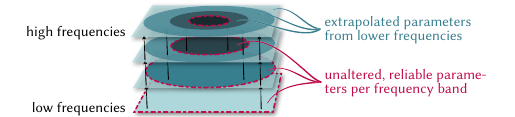}
    \setlength{\abovecaptionskip}{-4pt}
    \caption{Local extrapolation of reliable parameters across multiple scales towards higher frequencies.}
    \label{fig:extrapolation-scheme}
    \Description{}
\end{figure}

We aim to extract a set of local image statistics that can guide the recovery of missing frequency content: intensity, orientation, and phase. Given the foveation model, we know for each image location which spatial frequencies are preserved and which are attenuated. 
Therefore, we extract the parameters only from reliable frequency 
bands and extrapolate them to higher frequencies where content is missing (Figure \ref{fig:extrapolation-scheme}).

\begin{figure}
    \includegraphics[width=1\linewidth]{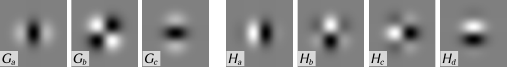}
    \setlength{\abovecaptionskip}{-4pt}
    \caption{Steerable quadrature basis filter set. The three $G$ and four $H$ filters form a steerable basis set each. If steered to an arbitrary rotation $\theta$, the resulting kernels $G_{\theta}$ and $H_{\theta}$ form a quadrature filter pair.}
    \label{fig:steerable-filters}
    \Description{}
\end{figure}

\paragraph{Filter Framework}
Each of the three extracted parameters presents distinct challenges and must be estimated as accurately and efficiently as possible. While intensity and orientation can, in principle, be estimated using simpler operators (e.g., Laplacian pyramids or gradient-based filters), doing so would require separate analysis pipelines per parameter. Phase estimation, however, fundamentally requires access to quadrature filter responses. To ensure consistency across parameters and scales, we therefore adopt a unified analysis framework, in particular, the steerable quadrature filter framework proposed by \citeauthor{freeman-1991a} [\citeyear{freeman-1991a}]. Their approach provides a compact basis of oriented band-pass filters from which responses at arbitrary orientations can be synthesized. The filter set consists of three even-symmetric filters $G$, corresponding to directional second derivatives of a Gaussian, and four odd-symmetric filters $H$, corresponding to a Hilbert transform of $G$ approximated by a third degree polynomial (Figure~\ref{fig:steerable-filters}). Together, these seven filters form a quadrature pair that enables the analysis of local phase, orientation, and magnitude. Filtering an input image with the seven kernels $G$ and $H$, yields seven base responses $G^a$--$G^c$ and $H^a$--$H^d$, which will be the only mean of analysis for the all parameters. Importantly, the response of an arbitrarily rotated kernel $G_{\theta}$ can be computed deterministically from the responses $G^a$--$G^c$. We note that the same basis filter set is later reused during synthesis, ensuring consistency between analysis and reconstruction (Section~\ref{sec:generating-noise-bands}).

\paragraph{Band-Pass Decomposition}
Prior to filtering, the image is decomposed into multiple levels using a Gaussian pyramid. This allows us to apply a fixed set of filters with constant spatial support at each level, resulting in a bandpass decomposition across scales. Since the synthesized signal is also band-limited, this representation enables frequency-specific reconstruction.

\paragraph{Color Handling}
Finally, analysis and reconstruction are performed solely on the luminance channel using a YCbCr color-luma decorrelation. Human visual sensitivity to chromatic detail is significantly lower than to luminance variations  \cite{ashraf-2024a}.
For computational reasons, we therefore focus on enhancing only the perceptually most prominent channel.

\subsection{Orientation}
We aim to recover a single predominant orientation per pixel. This choice is motivated by two considerations. First, the synthesis of phase-consistent structures across scales requires a unique local reference orientation. Second, at the spatial resolutions and eccentricities considered, perceptually relevant structure is typically dominated by a single orientation, while weaker secondary orientations contribute less to visual appearance.
To extract orientation information, we aim to find the orientation $\theta$ for which the response energy $E(\theta)= [G^{\theta}]^2 + [H^{\theta}]^2$ is maximal.
Following \citeauthor{freeman-1991a} [\citeyear{freeman-1991a}], $\theta$ can be retrieved deterministically with
\begin{equation}
    \theta = 0.5\arg(C_2, C_3)  \, ,
\end{equation}
where $C_2$ and $C_3$ is a linear combination of the responses $G^a$--$G^c$ and $H^a$--$H^d$ (see \cite{freeman-1991a} for definition).

\paragraph{Extrapolation}
The orientation field is extrapolated from coarser to finer scales using bilinear upsampling. To ensure correct circular interpolation, we represent the orientation $\theta$ as the phase of a complex number, upsample its real and imaginary components independently, and recover the orientation from the resulting phase.

\paragraph{Discussion}
Since no additional reliable orientation information is available at finer scales, this extrapolation yields approximately correct orientations while gradually losing fine directional detail as we upscale across multiple scales. Importantly, this procedure preserves orientation consistency across scales and allows us to maintain a coherent orientation–phase pair at each image location.

\subsection{Relative Phase and Cross-Scale Phase Alignment}

The phase field encodes the local alignment of neighboring filter responses and is critical for reconstructing coherent structures. Aligned phase across neighboring kernels results in well-defined edges, whereas random or misaligned phase produces jagged, incoherent structures (Figure~\ref{fig:importance-phase}). After determining the predominant orientation $\theta$ per pixel, we create locally steered responses $G^{\theta}$ and $H^{\theta}$ that form a complex pair $(G^\theta, H^\theta)$, from which the local phase is obtained as
\begin{equation}
    \phi = \arg(G^\theta, H^\theta)  \,.  
\end{equation}
Beyond the local phase alignment between neighboring kernels, phase correlation across scales is also critical for recovering coherent structures. Phase alignment across multiple scales encodes the geometric nature of local features: if even-symmetric filter responses align across scales, the result is a line-like structure, whereas alignment of odd-symmetric responses produces a step edge.

\paragraph{Extrapolation}
When extrapolating the phase field, we aim to separate its \textit{structural} content -- i.e., the phase relationships across scales -- from its spatial frequency content. This allows us to later reapply the structural relationships unchanged on top of an spatially upsampled phase field. We realize this by taking inspiration from a texture synthesis procedure from \citeauthor{portilla-2000a} [\citeyear{portilla-2000a}], who compute the relative phase $\Phi$ between two neighboring scales coarse $c$ and medium $m$ -- with $c$ and $m$ denoting the complex response $(G^\theta, H^\theta)$ of the respective scale each -- as
\begin{equation}
    \Phi(m,c) = \hat{c} \cdot m^*, \quad \hat{c} = \frac{c^2}{|c|}   \,.
\end{equation}
The information carried in $\arg(\Phi)$ determines how the responses of two filters across scales align, with $\pm\frac{\pi}{2}$ describing an edge, and $0$/$\pm\pi$ describing a line -- both of varying polarity. We extrapolate this specific behavior between $c$ and $m$ to a new, finer level $f$, by superimposing the across-scale behavior encoded in $\Phi$ to generate $\arg(f)$:
\begin{equation}
    \arg(f) = 2\arg(m) - \arg(\Phi)   \;.
\end{equation}
This is already a sufficient phase field for further synthesis. The magnitude $|f|$ will be defined in a separate step to ensure numerical stability (Section \ref{sec:intensity}).

\begin{figure}
    \includegraphics[width=\linewidth, page=1]{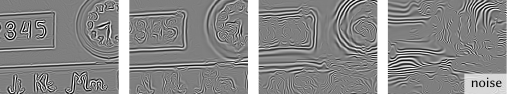}\\
    \vspace{0.05cm}
    \includegraphics[width=\linewidth, page=2]{fig/phase-upscaling.pdf}\\
    \vspace{0.05cm}
    \includegraphics[width=\linewidth, page=3]{fig/phase-upscaling.pdf}\\
    \vspace{0.05cm}
    \includegraphics[width=\linewidth, page=4]{fig/phase-upscaling.pdf} 
    \setlength{\abovecaptionskip}{-6pt}
    \caption{Examples of phase, magnitude and resulting noise if we upscale the parameters across multiple scales. The number of reconstructed lines increases, as the intensity and phase field lose their locality.
    \textcopyright \thinspace Christophe Seux}
    \label{fig:phase-extrapolation-flaws}
    \Description{}
\end{figure}

\paragraph{Discussion}
A particularity with this procedure is that the locality of structures will be scaled together with the spatial upsampling of $\Phi$. As a result, when phase information is extrapolated across multiple scales, the reconstructed lines lose their spatial precision and begin to appear duplicated (Figure~\ref{fig:phase-extrapolation-flaws}). While upsampling across a single scale preserves locality,  producing thin, well-defined lines, extrapolation across multiple scales results in a reconstruction that \textit{does} exhibit correctly aligned and clear lines,  however, multiple adjacent lines appear instead of a single one. This is not only due to the upsampling procedure of the phase, but also to the upsampling of the intensity maps $|f|$, which will be addressed in the following section. Whether the appearance of double lines is detrimental or beneficial for perceptual plausibility remains an open question for future work.

\subsection{Local and Global Intensity} \label{sec:intensity}

The intensity field controls the local energy of the synthesized bandpass content, as visible in Figure \ref{fig:phase-extrapolation-flaws}. An important requirement for our content synthesis is that untextured regions of approximately constant luminance, such as a clear sky, remain untextured after enhancement. Consequently, intensity values must be extrapolated in a way that preserves this behavior across frequency bands.

\paragraph{Extrapolation}
\citeauthor{tariq-2022a} [\citeyear{tariq-2022a}] base their intensity extrapolation on the assumption that neighboring frequency bands exhibit similar energy. In our framework, however, intensity parameters may need to be extrapolated across multiple scales. Naïvely propagating intensity from naturally over-represented low frequencies to higher frequency levels would lead to an over-representation of high frequencies, and ultimately an unnatural, flattened power distribution. To address this, we apply a global intensity adjustment $a_{\sigma}$ as we extrapolate intensity across scales. We base this adjustment on the observation that natural images exhibit typically a linear power spectrum of slope $\approx2$ \cite{reinhard-2009a}. We estimate the specific slope of the present image based on the preserved frequencies, and extrapolate this information to retrieve $a_{\sigma}$ per recovered level. The resulting intensity estimate $\sigma$ for a given scale is therefore
\begin{equation}
    \sigma =  a_{\sigma}  |(G^\theta , H^\theta)|  \; ,
\end{equation} 
where $|(G^\theta , H^\theta)|$ is the phase-invariant magnitude of the nearest reliable scale, quantified by the magnitude of the steered complex quadrature response. Detailed reasoning on the retrieval of $a_{\sigma}$ is provided in the Supplementary Materials.

\paragraph{Discussion}
This extrapolation procedure preserves local intensity across scales without introducing additional structure. However, this procedure is also the true reason behind the double edges appearing in Figure \ref{fig:phase-extrapolation-flaws}. Although one could consider thinning or eroding the lines that appear in the magnitude field after upscaling, doing so deterministically is not feasible for complex stimuli.
\section{Synthesis of Missing Frequency Content} \label{sec:generating-noise-bands}

We aim to synthesize the missing frequency content as a locally oriented, phase shifted and intensity adjusted bandpass signal. For this, we follow a modified Gabor-noise procedure \cite{tavernier-2019a}. We generate an impulse map, of which each impulse is convolved with a steered, phase-shifted and intensity adjusted kernel from our steerable filter set. Each kernel's properties are controlled by the three parameters denoted at the impulse's location. The kernel $K$ at a given location is therefore defined as
\begin{equation}\label{eq:kernel}
K \; = \; \left[ \cos(\phi)G_\theta - \sin(\phi)H_\theta \right] \cdot \sigma \;.
\end{equation}
As we already computed the steered responses $G^{\theta}$ and $H^{\theta}$, one might tend to directly phase-shift these images themselves to acquire the correct frequency layer by performing $\cos(\phi)G^\theta - \sin(\phi)H^\theta$ for the full image at once. However, as the parameter fields are not bandpass limited, this pixel-wise procedure would introduce frequencies outside of the desired band. The resulting signal will contain all frequencies, which is undesirable. Contrarily, applying the parameters 
to the full kernel ensures that the spectral response of the synthesized content remains a correct bandpass per scale.

During synthesis, we ensure an even density of impulses across the image. However, as the kernel density increases, the global variance and therefore global intensity of the synthesized content increases too. To control for this, we match the global standard deviation of the content's histogram with our estimated $a_{\sigma}$. This follows the same reasoning as mentioned in Section~\ref{sec:intensity}. In Figure \ref{fig:frequency-bands-histogram}, the bandpass nature of the generated layers is visible. The highest band of frequencies will not be reconstructed by our method, as our kernel is limited to a maximum frequency of 0.25\thinspace \gls{cpp}.

\paragraph{Acceleration}
In practice, we do not perform any convolution at runtime. Instead, we subdivide the impulses into several sub-maps, such that the kernels will not overlap after convolution. Next, we convolve each sub-map once with each basis filter and store it. Doing this allows us to recover an arbitrary response per impulse location by linearly combining the seven pre-convolved maps during runtime. By efficiently handling the impulse locations and parameter maps, we solely rely on a single set of seven pre-convolved images. The linear combination of which can be computed in parallel on the GPU, making our synthesis procedure very fast. A detailed outline of the required operations per frame is listed in the Supplementary Material.
\section{Implementation Details}

\begin{figure}
\begin{minipage}{122.68 pt}
    \centering
    \includegraphics[page=1]{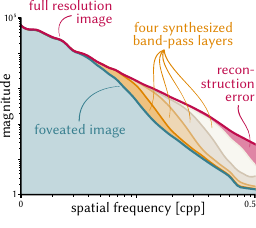}
    \setlength{\abovecaptionskip}{-8pt}
    \captionof{figure}{Exemplary frequency histogram of the \textsc{Classroom} scene after foveation and enhancement.}
    \label{fig:frequency-bands-histogram}
    \Description{}
\end{minipage}%
\hfill%
\begin{minipage}{97.62 pt}
    \centering
    \includegraphics[page=2]{fig/frequency-histogram_benchmark.pdf}
    \setlength{\abovecaptionskip}{-8pt}
    \captionof{figure}{Computational times of our post-processing pass for various image resolutions.}
    \label{fig:benchmark}
    \Description{}
\end{minipage}
\end{figure}

We implement our method both as a Python prototype and as a real-time system in Unity 6000.0.9. The straightforward real-time implementation is based on a custom render pass which is injected as a post-processing step.
Performance is evaluated for image resolutions up to 4K, with its execution times for the individual pipeline stages shown in Figure~\ref{fig:benchmark}.
The analysis times are broken down into 
the setup steps, such as color conversion and computation of the eccentricity map, computation of the image pyramids, filtering of the image with $G$/$H$, and the retrieval of $\sigma$, $\theta$ and $\phi$. The synthesized content is generated with one impulse per pixel, resulting in the maximum achievable spatial quality.
All measurements are obtained using the Unity Profiler in play mode on a system equipped with an NVIDIA GeForce RTX 4090 and Intel Core i9-13900KF CPU.
Further acceleration could be achieved by synthesizing the content with a lower impulse density, or by limiting the synthesis resolution in peripheral image regions, where high spatial frequencies become imperceptible due to reduced visual sensitivity, allowing unnecessary 4K- and even 2k-frequency detail to be omitted.

As our method operates purely as a post-processing step, it assumes an input that is already spatially and temporally antialiased. While this preprocessing introduces additional cost, it can be offset by reducing the initial sampling rate. In particular, \citeauthor{guenter-2012a}~\citeyear{guenter-2012a} showed that rendering at lower sampling rates combined with antialiasing is overall more efficient than densely sampling the image without such reconstruction. Moreover, assuming an antialiased input is standard practice for high-quality rendering and recent foveated rendering methods \cite{patney-2016a,tursun-2019a,tariq-2022a}.

\section{Perceptual Calibration Study}\label{sec:calibration-study}

\begin{figure}
    \includegraphics[width=\linewidth]{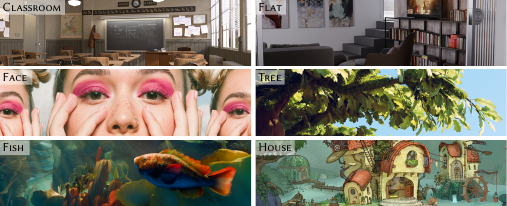}
    \setlength{\abovecaptionskip}{-4pt}
    \caption{Images used in the perceptual study.
    \textcopyright \thinspace Christophe Seux, Flavio Della Tommasa, Wei Ding, Simon Thommes/Blender Studio, and Blender Studio}
    \label{fig:study-images-overview}
    \Description{}
\end{figure}

While our method aims to recover local image statistics as accurately as possible, the relative perceptual impact of each recovered property remains unclear. We therefore conduct a study to evaluate the perceptual importance of each recovered parameter and to quantify the magnitude of its contribution by progressively enabling subsets of the enhancement. We aim to determine the range of high spatial frequencies which can be substituted by our synthesized content, without participants being able to discern a difference in image quality compared to the full resolution image. We test our enhancement procedure directly on a variety of complex stimuli of various kind and style (Figure~\ref{fig:study-images-overview}). This allows for direct applicability of our determined thresholds for real-world use cases, contrary to thresholds evaluated on simple stimuli (vision science approach), which might suffer from confounding factors when being applied to natural stimuli.

\begin{figure*}
    \includegraphics[width=1\linewidth]{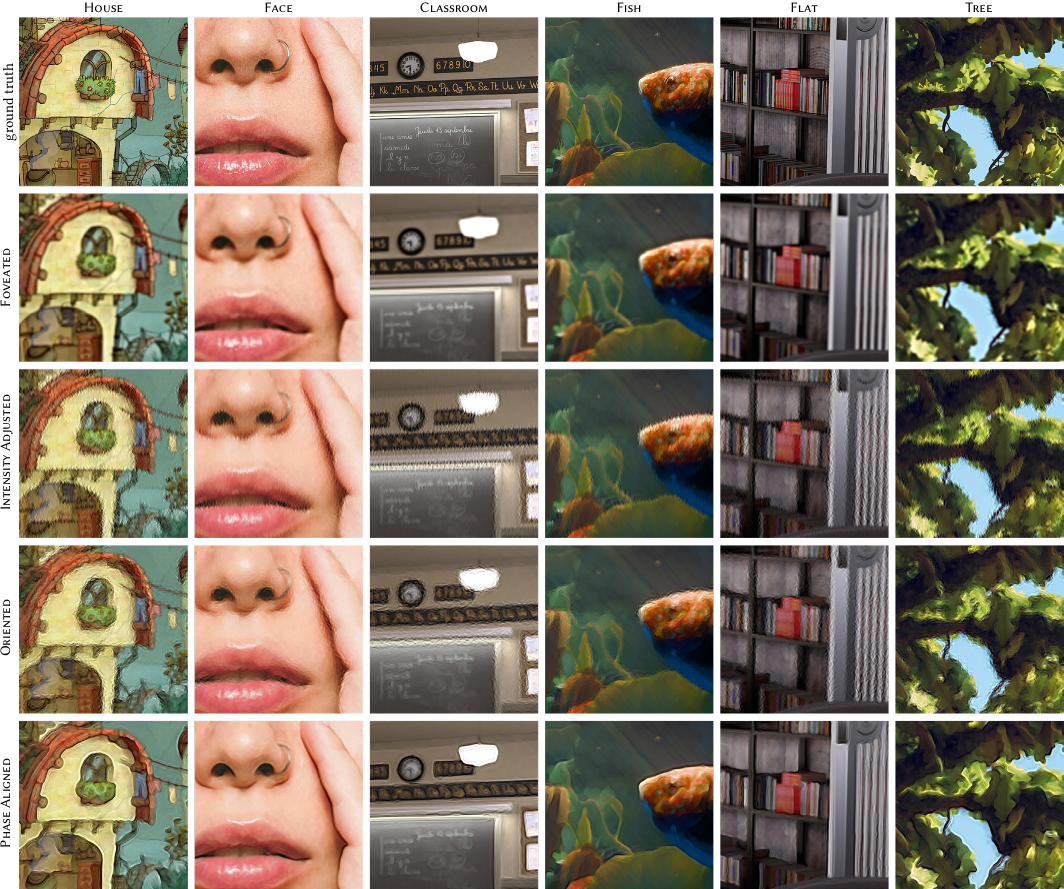}
    \caption{Patches from the modified images used in the user study. The base images were downsampled to a shading rate of 1/64, and parameters were extrapolated across three scales. In the \textsc{Phase Aligned} case, edges get reconstructed much more clearly, compared to the other cases.}
    \label{fig:image-collection}
    \Description{}
\end{figure*}

\begin{figure*}
    \includegraphics[width=1\linewidth]{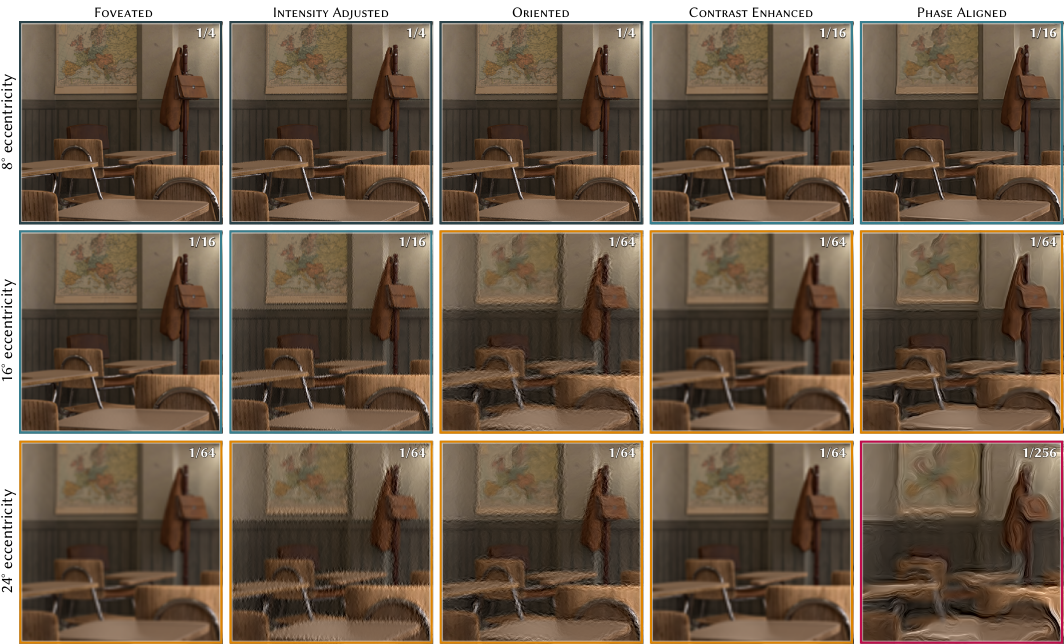}
    \caption{Exemplary images which are foveated and enhanced with various methods examined in the calibration study. All displayed images in one row were indistinguishable to the full resolution reference at the indicated eccentricity. The small numbers indicate the shading rate of the input image.}
    \label{fig:classroom-grid}
    \Description{}
\end{figure*}

\paragraph{Independent Variables}
We control three independent variables: enhancement method, shading rate of the input image, and eccentricity. Regarding the enhancement method, we examine the following five quality grades: 
\textsc{Foveated}, being the unmodified input image of reduced shading rate (upsampled Gaussian pyramid layer);
\textsc{Contrast Enhanced}, being the input image with added contrast enhancement following \citeauthor{patney-2016a}~\shortcite{patney-2016a};
\textsc{Intensity Adjusted}, for which we adjust the enhancement layer solely considering intensity $\sigma$;
\textsc{Oriented}, recovering additionally $\theta$;
and \textsc{Phase Aligned}, recovering the full set of parameters.
Exemplary images are displayed in Figure~\ref{fig:teaser}, \ref{fig:image-collection} and \ref{fig:classroom-grid}. Our method recovers full layers of an image pyramid, so the investigated frequency bands are spaced logarithmically. During testing, we replace a varying amount of the high frequency bands with the synthesized content. The amount of replaced frequencies is determined by the shading rate of the underlying input image. To examine the participants' sensitivity at various locations in the visual field, we apply foveal masking of $8^\circ$, $16^\circ$, or $24^\circ$. To examine sensitivity at the fovea ($0^\circ$), participants are allowed to freely look at the images without constraints. During the procedure, we present each condition (enhancement method $\times$ shading rate $\times$ eccentricity) once per scene, resulting in six measurements per condition and participant. This ensures that our estimated data holds reliably for a variety of natural stimuli.

\begin{figure}
    \includegraphics[width=\linewidth]{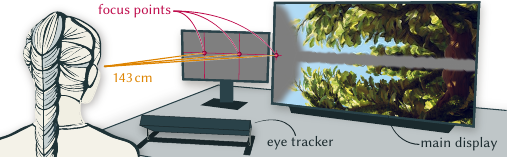}
    \setlength{\abovecaptionskip}{-4pt}
    \caption{Study setup used for the Calibration Study. The same setup was mirrored on the right side of the central display.
    \textcopyright \thinspace Simon Thomas}
    \label{fig:setup-displays}
    \Description{}
\end{figure}

\paragraph{Study Setup and Apparatus}
The experiment was conducted on a 55' LG OLED screen with a native resolution of $3480\times2160$ pixels, flanked by two helper displays positioned on either side (Figure~\ref{fig:setup-displays}). Participants were seated at a viewing distance of 143\thinspace cm from the main display, resulting in a resolution of  \textasciitilde40\thinspace\gls{cpd} when looking straight ahead. Participants were seated in front of the left or right edge of the main display to accommodate high eccentricities without excessive masking of the foveal masking. The stimuli were distributed randomly across both sides. Their gaze was monitored using a Tobii Spectrum Pro eye tracker. Head movements were restricted using a chin rest to ensure stable viewing conditions. The experimental procedure was implemented in Unity 2022.3.9. All stimuli were pre-generated using Python scripts and presented without modification during runtime.

\paragraph{Procedure and Task}
The experiment consisted of an initial free viewing phase, followed by peripheral viewing under fixation. In the first phase, participants were seated centered in front of the display and observed all stimuli freely without gaze and time constraints. This phase served both to familiarize participants with the full resolution scenes, and to expose them to the visual artifacts produced by the different enhancement methods. Each stimulus presentation consisted of a horizontally mirrored display of two images, of which one was the full-quality reference image and the other was our reconstructed version (Figure~\ref{fig:setup-displays}). Participants’ task was to determine which of the two images is the unmodified full resolution image (2AFC procedure). For stimuli presented at higher eccentricities, the presentation time was limited to 1.5\thinspace s to prevent visual artifacts such as afterimages \cite{wallace-2013}.

\paragraph{Participants}
We conducted the study on 11 voluntary participants (5\thinspace F, 6\thinspace M) with an average age of $26\pm3$ years. All participants had normal or corrected to normal vision. Six participants reported prior experience with perceptual experiments. All participants signed previous consent, and were compensated for their time. The study was approved by the institutional ethics committee.

\subsection{Results}\label{sec:calibration-results}

\begin{figure*}[t]
    \includegraphics[width=\linewidth]{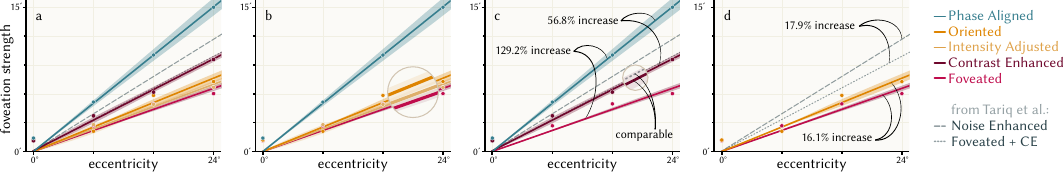} 
    \setlength{\abovecaptionskip}{-4pt}
    \caption{
    Estimated 75\% thresholds and respective linear fits per enhancement method.
    The width of the line denotes the uncertainty of the estimated 75\% thresholds.
    The y-axis describes the $\sigma$ of a Gaussian kernel with which the original image can be convolved, while remaining at indistinguishable visual quality after enhancement. CE abbreviates contrast enhancement.
    (a) Combined plots, 
    (b)--(d) show breakout plots from (a) for readability.}
    \label{fig:results-fit}
    \Description{}
\end{figure*}

We collect a total of 7556 measurements, equaling 66 individual measurements per condition.
We aim to find the lowest shading rate necessary per enhancement method $\times$ eccentricity pair, for which we could substitute high frequency content with our reconstruction, without participants being able to tell which of the presented images is the unaltered one. For this, we fit a psychometric function to the data and determine the 75\% detectability point as the threshold, from which on participants could clearly discern the modified and the unmodified image. Plots of the data, fitted functions, and further details can be found in the Supplementary Material. Next, we model the behavior of the thresholds per enhancement method across eccentricities by fitting linear functions (Figure \ref{fig:results-fit}).

\paragraph{Impact of the Number of Considered Parameters}
Figure~\ref{fig:results-fit}b shows the effect of progressively recovering additional parameters. All displayed methods do not apply contrast enhancement, and therefore solely display the effect of our enhancement in isolation. As expected, the tolerated foveation strength increases as more parameters are recovered.

\paragraph{Impact of Kernel Choice}
We observe surprisingly dissimilar increases between the \textsc{Foveated}, \textsc{Oriented}, and \textsc{Phase Aligned} conditions. We hypothesize that the perceptual relevance of correct orientation is strongly influenced by the choice of synthesis kernel. In our implementation, the kernel exhibits only weak orientational selectivity (Figure~\ref{fig:steerable-filters}). As a result, orienting the kernel improves structural coherence to some extent, but the overall visual impact remains limited. In contrast, kernels with stronger anisotropy would be expected to exhibit a much larger perceptual difference between correct and incorrect orientations, as orientation errors would become more visually salient. Importantly, the relationship between kernel shape and perceptual impact differs for phase alignment. Here, the opposite trend may apply: smaller and more spatially localized kernels allow for more precise reconstruction of fine-scale structures. We hypothesize that this is the reason for the dissimilar improvements between our \textsc{Phase Aligned} and \textsc{Oriented} conditions.

\paragraph{Impact of Contrast Enhancement}
Figure~\ref{fig:results-fit}d compares the performance of our method against the averaged results reported by \citeauthor{tariq-2022a} [\citeyear{tariq-2022a}]. While the absolute thresholds achieved by our method are lower overall, the relative effect of adding oriented enhancement to a foveated base image is comparable. This indicates that our method reproduces previously observed perceptual trends despite differences in absolute scaling. Moreover, the foveation thresholds measured for the contrast enhanced baseline alone closely match those reported by \citeauthor{tariq-2022a} [\citeyear{tariq-2022a}] (Figure \ref{fig:results-fit}c), further indicating that our experimental setup yields directly comparable results. We therefore attribute the absolute difference to the absence of contrast enhancement in our method.
However, it remains unclear whether combining contrast enhancement with our \textsc{Phase Aligned} synthesis would too further improve performance. Nevertheless, the overall trend is unambiguous: aligning phase yields a substantial improvement for our method. With our current kernel choice, we observe a +129\% increase between the \textsc{Foveated} and \textsc{Phase Aligned} conditions, and still  a +57\% increase when contrast enhancement is enabled. This outperforms the framework of \citeauthor{tariq-2022a}~\shortcite{tariq-2022a} by 39\% in direct comparison.
\section{Validation Experiment}

Finally, we aim to verify that our previous findings hold for interactive scenarios. To this end, we conduct a validation study in which participants freely observe fully foveated and post-enhanced images. For each stimulus, we compare our enhancement method (using the \textsc{Phase Aligned} framework) against the method of \citeauthor{tariq-2022a} [\citeyear{tariq-2022a}], which represents the closest and most recent state-of-the-art competitor. While additional foveated rendering techniques exist, many of them address orthogonal stages of the foveated rendering pipeline, such as sampling \cite{tursun-2019a,krajancich-2023a} or antialiasing \cite{patney-2016a,kaplanyan-2019,guenter-2012a}. Our technique operates solely in the final step of the pipeline, which is enhancement of the rendered image. We therefore focus our comparison on the most closely related method that also operates at the same stage \cite{tariq-2022a}.

\paragraph{Stimuli}
Stimuli are generated as follows: 
First, full-resolution images are foveated according to the foveation model shown in Figure~\ref{fig:results-fit}, using our most competitive configuration (\textsc{Phase Aligned}).
Next, two enhancement layers are synthesized: one using the procedure of \citeauthor{tariq-2022a} [\citeyear{tariq-2022a}], and one using our proposed framework.
In both cases, the enhancement is added solely to the luminance channel of the foveated image. In a second condition, contrast enhancement is additionally applied to both versions using the parameters reported by \citeauthor{tariq-2022a} [\citeyear{tariq-2022a}]. This procedure results in a total of 12 comparison pairs (six scenes $\times$ with/without contrast enhancement).

\paragraph{Task}
Participants sequentially observe pairs of enhanced images alongside the full-resolution reference image. Their task is to select the image that appears closer to the reference, following a 2AFC procedure. Each comparison pair is constructed such that both enhancement methods either include or exclude contrast enhancement. This ensures that the comparison isolates the effect of the high-frequency enhancement.

\paragraph{Participants}
We collect data of 15 voluntary participants (6\thinspace F, 9\thinspace M), with an average age of $27\pm3$ years. All participants had normal or corrected to normal vision, and gave previous consent. The study was approved by the institutional ethics committee.

\subsection{Results}

\begin{figure}
    \includegraphics[page=2]{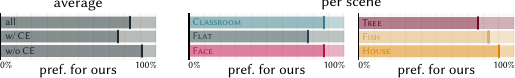}
    \setlength{\abovecaptionskip}{-4pt}
    \caption{Results of the Validation Study. The values report the percentage of cases in which participants favored our method. All displayed results are significant ($p<0.05$).}
    \label{fig:results-validation-study}
    \Description{}
\end{figure}

Figure~\ref{fig:results-validation-study} shows the average percentage of participants who perceived our method as closer to the full-resolution reference. 
Overall, 
our approach is preferred in 83\% of the comparisons ($p=0.035$). A breakdown by the presence/absence of contrast enhancement reveals an even stronger advantage for our method when contrast enhancement is absent, achieving a preference rate of 91\% ($p<0.001$). Importantly, the contrast enhancement used in the method of \citeauthor{tariq-2022a} \shortcite{tariq-2022a} was explicitly tuned to optimally complement their noise synthesis. Even under these favorable conditions, our method still achieves a preference rate of 75\% in the contrast-enhanced case. This demonstrates that our approach outperforms the procedure of \citeauthor{tariq-2022a}~\shortcite{tariq-2022a} even when the baseline is optimized for their method. 
We note that our evaluation employs foveation strengths that exceed the intensity range originally considered by \citeauthor{tariq-2022a} [\citeyear{tariq-2022a}]. At these stronger foveation levels, degradation of their method is expected. In contrast, our approach is explicitly tuned to handle such foveation strengths in the non–contrast-enhanced setting, which is reflected in the strong preference observed in this condition. For the contrast-enhanced case, our method was not specifically optimized for these extreme settings; nevertheless, it still achieves a preference rate of 75\%. These results suggest that our method remains competitive beyond the intended operating range of the baseline, while being particularly robust in the non–contrast-enhanced case.

\section{Limitations and Future Work}

Our current approach has several limitations that suggest directions for future work.
First, our method operates on discrete pyramid levels. As a result, the decision of which frequencies are rendered versus synthesized is quantized to entire pyramid bands. A more desirable formulation would allow a continuous control over the frequency threshold between rendered and synthesized content, enabling finer trade-offs between rendering cost and perceptual quality.
Second, while phase alignment enables the reconstruction of coherent structures, fine edges cannot be perfectly recovered. When phase information is extrapolated across multiple scales, edges tend to multiply, leading to several closely spaced lines instead of a single crisp boundary. As a result, reconstructed structures retain their semantic identity (e.g., lines remain lines) but lack precise localization (Figure~\ref{fig:phase-extrapolation-flaws}).
Third, the synthesis is constrained by the choice of kernel. For computational reasons, our current kernels are spatially small and operate at frequencies defined by the image pyramid, limiting both frequency resolution and the ability to finely control spectral placement.
Finally, we did not explicitly evaluate the temporal behavior of the synthesized content. Our method relies primarily on simple filtering operations, which are expected to behave smoothly under gradual input changes. However, some intermediate estimates, such as local orientation of fine spatial structures, may still vary abruptly in dynamic scenes. Prior work mitigates such artifacts by restricting parameter estimation to coarser levels of the Gaussian pyramid \cite{tariq-2022a}. However, this approach suppresses high-frequency details and is therefore not compatible with our method. Addressing temporal artifacts in video thus remains an open direction for future work.
\section{Conclusion}

As virtual and augmented reality technologies and devices continue to develop, rendering efficiency requirements will further increase. Foveated rendering remains a vital strategy for balancing increasing display quality with the computational demands it entails. In this context, understanding the requirements of human perception and aligning them with rendering methods to create inexpensive metamers remains a challenging yet fascinating research topic. Inspired by prior work on image statistics, this paper examines the role of phase information in generating peripheral metamers. We demonstrate that phase information estimated from a foveated image can be used to improve its quality and, in turn, enable the creation of more efficient metamers. We present perceptual experiments supporting these claims, as well as a new method for synthesizing such metamers. While future work may investigate even more image statistics to improve the recovery of omitted spatial detail, a broader and more fundamental question arises: What information in the input is truly necessary for efficient synthesis of such visual metamers?


\bibliographystyle{ACM-Reference-Format}
\bibliography{reference}

\end{document}


\title{It’s Not Just a Phase: Creating Phase-Aligned Peripheral Metamers}
\subtitle{Supplementary Material}

\begin{teaserfigure}
   \vspace{0.5cm}
\end{teaserfigure}

\maketitle

\section{Global Intensity Tuning of the Frequency Bands}\label{sec:global-intensity-adjustment}

We estimate the global intensity of one synthesized bandpass signal by tuning its standard deviation to match the standard deviation which we would expect the respective Laplace layer to have. For this, we follow the intuition that
the slope of the power spectrum of natural images is mostly linear with a slope of $\approx 2$ \cite{reinhard-2009a}.
According to Parseval's theorem \cite{parseval-1806a}, we can assume that the sum of the square of a function is equal to the sum of the square of its Fourier transform:
$$
\sum_{x,y}  f(x,y)^2
\;=\;
\frac{1}{M^2}\sum_{u,v} \lvert F(u,v)\rvert^2
\;=\;
\sum_{u,v}  S(u,v),
$$
with $f$ being the image in the spatial domain, $F$ is the Fourier transform of $f$, and $S$ is the power spectrum.
In our case, $f$ is an individual Laplace layer in the spatial domain. The squared sum of the Laplacian pyramid layers should therefore reflect a bandpass analysis of the power spectrum. 
Further, the expected value of the power spectrum $\mathbb{E}[S]$ is equal to the expected value of the square of the image in the spatial domain, $\mathbb{E}[f^2]$. 
Further, we can define the variance as $\text{Var}(f) = \mathbb{E}[f^2] - \mathbb{E}[f]^2$. The term $\mathbb{E}[f]^2$ essentially denotes the squared mean on the pixel values. For our bandpass decomposition, we can assume that the mean should be close to 0, and, therefore, disregard this term.
From here, we infer that the mean spectral power is directly related to the variance in the spatial domain:
$$\text{Var}(f) \; \approx \; 
\mathbb{E}[f^2] \; := \; 
\frac{1}{M^2} \sum_{x,y}f(x,y)^2  \; =\;  
\frac{1}{M^2} \sum_{u,v}  S(u,v).$$
Based on this, we can simply linearly extrapolate the behavior of $\mathbb{E}[f^2]$ across multiple frequency levels to arrive at a good assumption for the strength of our newly created content.

\section{Computational of Analysis/Synthesis}
\paragraph{Parameter Analysis}
The input image is transposed to the YCbCr color model.
Next, we construct a Gaussian and Laplacian pyramid. Depending on the applied foveation model, and therefore the severity with which high frequencies get attenuated, the required depth of the Gaussian pyramid varies. However, for the computation of proper global intensity, a deep pyramid is preferable.
Each Gaussian pyramid layer is convolved using the seven basis filters. All filters are separable, resulting in 14 $1\times9$ convolutions per layer.

\paragraph{Parameter Extrapolation}
Each extrapolation step consists of an upscaling process, partially paired with linear combinations.
There is no filtering involved in this step.

\paragraph{Synthesis}
To synthesize the missing signals, we prepare seven pre-filtered impulse maps prior to starting the real-time application, and keep them in the working memory of the GPU.
During runtime, we subsample the parameter fields at the respective impulse locations, and upscale them again in a nearest neighbor fashion. This ensures that one of the pre-convolved kernels is completely covered by a single value in the parameter map. Once we ensure this, we can use simple linear combinations of the modified parameter fields and the pre-convolved maps. Of course, we have to repeat this procedure for each of the impulse sub-sets. Having 1 impulse per $2 \times 2$ region, we perform this procedure 16 times. Having 1 impulse per pixel requires 64 of these operations. The operations are parallelizable on the GPU.

\paragraph{Code Release}
Our full implementation will be released on GitHub after publication of the manuscript.

\section{Results Plots and Analysis}

\begin{figure*}
    \includegraphics[width=\linewidth]{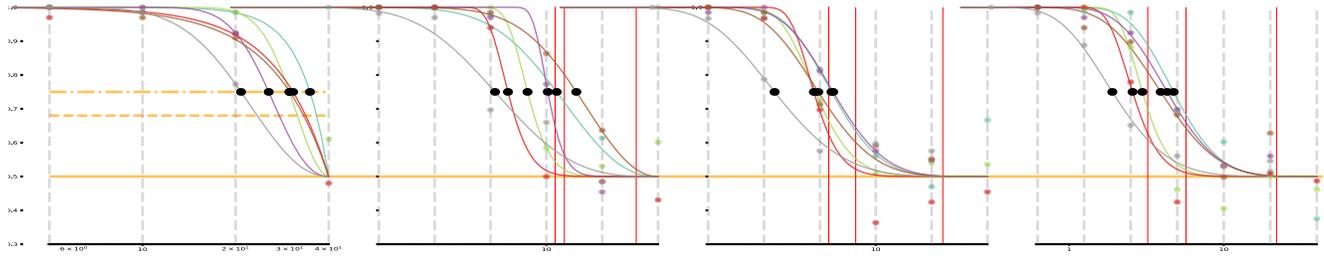}
    \caption{Individual detection percentages per enhancement method, eccentricity and level. The dashed vertical lines indicate the Gaussian pyramid layer of the base image. One step corresponds therefore to a shading rate reduction of factor 4.}
    \label{fig:fits}
\end{figure*}

We fit a psychometric function (Weibull curve) which is anchored at 50\% detection probability for 1/1 shading rate (no foveation) and growing to 100\% towards lower shading rates. Figure \ref{fig:fits} shows the fit, as well as the retrieved 75\% threshold per eccentricity and enhancement method.

\bibliographystyle{ACM-Reference-Format}
\bibliography{reference}